\documentclass[preprint,longtitle,authoryear]{elsarticle}

\journal{arXiv}

\usepackage{latexsym}
\usepackage{amsmath}
\usepackage{amssymb}
\usepackage{algorithm}
\usepackage{algorithmic}
\usepackage{hyperref}
\usepackage{graphicx}

\def\O{\mathcal{O}}
\def\c{{\rm cl}}

\begin{document} 

\begin{frontmatter}

\title{{\bf A Concise and Formal Definition of RAF Sets and the RAF Algorithm}}

\author{Wim Hordijk}
\address{SmartAnalytiX, Vienna, Austria\\
{\tt wim@WorldWideWanderings.net}}

\begin{abstract}
Autocatalytic sets are self-catalyzing and self-sustaining chemical reaction networks that are believed to have played an important role in the origin of life. They have been studied extensively both theoretically as well as experimentally. This short note provides (1) a complete and formal definition of autocatalytic sets (or RAF sets), and (2) an efficient algorithm to detect such sets in arbitrary reaction networks. Although both have been presented in various forms in earlier publications, this note serves as a concise and convenient reference.
\end{abstract}

\begin{keyword}
Autocatalytic sets \sep RAFs \sep origin of life \sep chemical reaction networks
\end{keyword}

\end{frontmatter}

\section{Introduction}

An autocatalytic set is a chemical reaction network in which each reaction is catalyzed by at least one of the molecules involved in the network, and where all these molecules can be produced through a sequence of reactions from the network itself. In other words, an autocatalytic set is a {\it self-catalyzing} and {\it self-sustaining} chemical reaction network.

The original concept of autocatalytic sets was introduced by \citet{Kauffman:71,Kauffman:86,Kauffman:93}, and has been studied extensively since then, both theoretically as well as experimentally \citep{Hordijk:19}. It was mathematically formalized as {\it reflexively autocatalytic and food-generated} sets, or RAF sets \citep{Steel:00,Hordijk:04,Hordijk:15,Hordijk:17}. Autocatalytic sets are believed to have played an important role in the origin of metabolism and life \citep{Hordijk:10,Xavier:20,Hordijk:22}.

Although many research articles have been published about autocatalytic sets in general and RAF sets more specifically \citep{Hordijk:19}, most of these articles emphasized certain aspects more than others, depending on the specific results presented or the audience intended. As a consequence, the fundamental mathematical details have become somewhat scattered throughout the literature. With an increasing interest in autocatalytic sets, the demand for a concise and clear description is also growing. Therefore, this short note serves as a convenient reference, providing a complete and formal definition of RAF sets and of the RAF algorithm for detecting them in arbitrary reaction networks.

\section{RAF sets}

First, we define a {\it chemical reaction system} (CRS) as a tuple $Q = \{X,R,C,F\}$, where:
\begin{itemize}
\item $X = \{x_1, x_2, \ldots, x_n\}$ is a set of molecule types.
\item $R = \{r_1, r_2, \ldots, r_m\}$ is a set of reactions. A reaction $r$ is an ordered pair $r=(A,B)$ where $A, B \subset X$. The (multi)set $A = \{a_1, \ldots a_s\}$ are the reactants and the (multi)set $B= \{b_1, \ldots, b_t\}$ are the products.
\item $C \subseteq X \times R$ is a set of catalysis assignments. A catalysis assignment is a pair $(x,r)$ with $x \in X$ and $r \in R$, indicating that molecule type $x$ can catalyze reaction $r$.
\item $F \subset X$ is a food set, i.e., molecule types that can be assumed to be available from the environment.
\end{itemize}

Given a CRS $Q$, a subset $R'$ of $R$, and a subset $X'$ of $X$, we define the {\it closure} of $X'$ relative to $R'$, denoted $\c_{R'}(X')$, to be the (unique) minimal subset $W$ of $X$ that contains $X'$ and that satisfies the condition that, for each reaction $r=(A,B)$ in $R'$,
\[ A \subseteq  X' \cup W \Longrightarrow B \subseteq W. \]
Informally, $\c_{R'}(X')$ is $X'$ together with all molecules that can be constructed from
$X'$ by the repeated application of reactions from $R'$ (regardless of whether a reaction is catalyzed or not).

Now, given a CRS $Q=\{X,R,C,F\}$ and a subset $R'$ of $R$, $R'$ is:
\begin{enumerate}
\item {\it reflexively autocatalytic} if for each reaction $r \in R'$, there is at least one molecule in the closure of the food set (relative to $R'$) that can catalyze reaction $r$, and
\item {\it food-generated} if all reactants of all reactions $r \in R'$ are in the closure of the food set (relative to $R'$).
\end{enumerate}
A subset of reactions $R'$ is {\it reflexively autocatalytic and food-generated} (or RAF) if both conditions hold \citep{Hordijk:04}.

More formally, given a CRS $Q=\{X,R,C,F\}$ and a subset $R'$ of $R$, $R'$ is a RAF set if for each $r=(A,B) \in R'$:
\begin{enumerate}
\item $\exists x \in \c_{R'}(F): (x,r) \in C$, and
\item $A \subseteq \c_{R'}(F)$.
\end{enumerate}
In words, a subset of reactions $R'$ is a RAF set if for each of its reactions at least one catalyst and all reactants are in the closure of the food set (relative to $R'$).

\begin{figure}[H]
\centering
\includegraphics[scale=0.45]{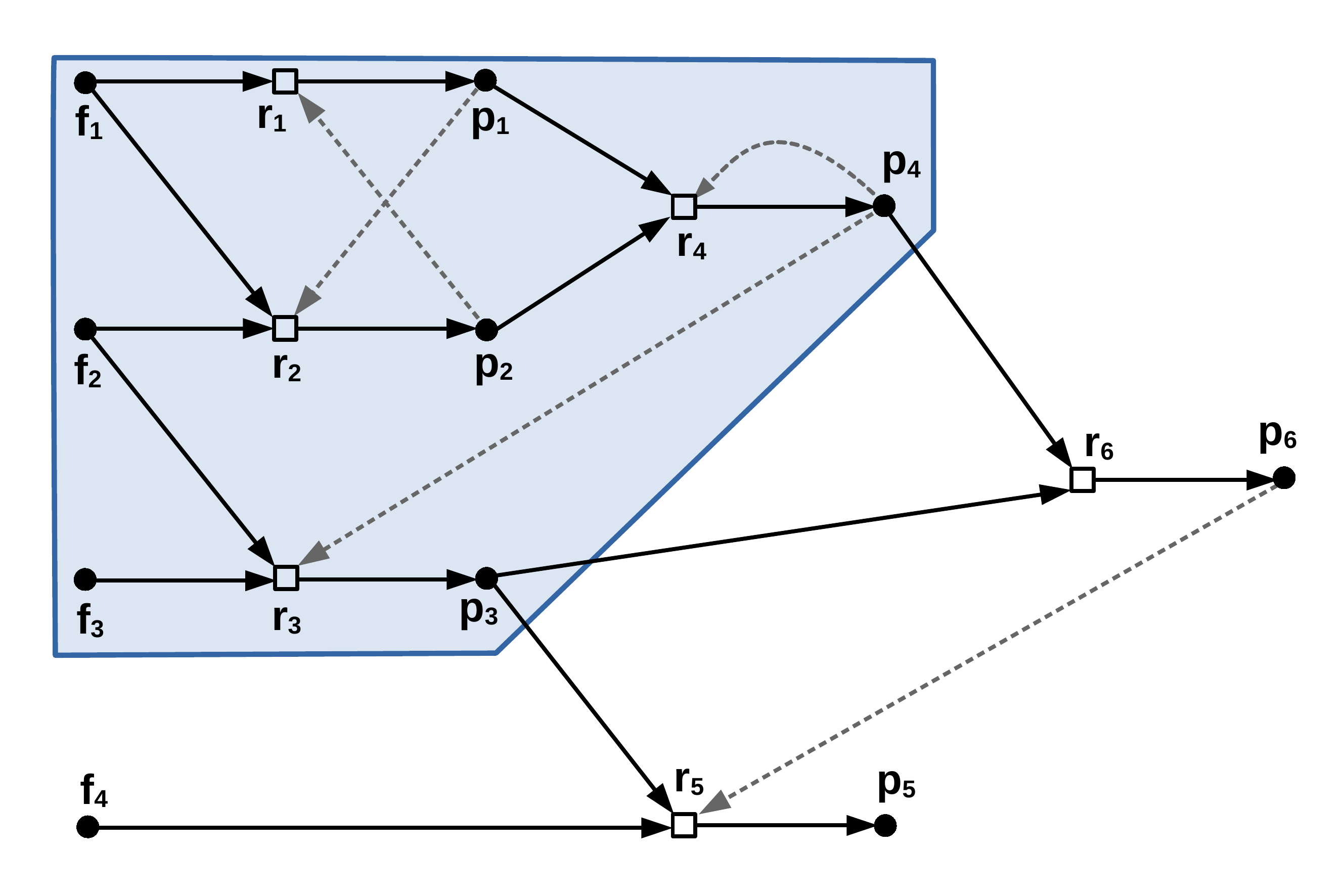}
\caption{An example chemical reaction network containing a RAF set (blue outline).}
\label{fig:example}
\end{figure}

Fig. \ref{fig:example} shows an example with $X = \{f_1,\ldots,f_4,p_1,\ldots,p_6\}$, $F = \{f_1,\ldots,f_4\}$, $R=\{r_1,\ldots,r_6\}$, and $C = \{(p_1,r_2), (p_2,r_1), (p_4,r_3), (p_4,r_4), (p_6,r_5)\}$. The graph representation used is common for chemical reaction networks, with dots denoting molecule types and boxes denoting chemical reactions \citep{Temkin:96}, and with catalysis added as dashed grey arrows.

The subset $R'=\{r_1,r_2,r_3,r_4\}$, contained within the blue outline, is a RAF set within the full network. Note that reaction $r_6$ is not catalyzed at all, and can thus not be part of a RAF set. Furthermore, this reaction produces molecule type $p_6$, which is the only possible catalyst for reaction $r_5$, which can thus also not be part of a RAF set.

Finally, the more restricted notion of a {\it constructively autocatalytic and food-generated} set (or CAF) is a RAF $R'$ with the additional condition that the reactions in $R'$ can be ordered in such a way that for each reaction $r_i \in R'$ in this ordering, each reactant and at least one catalyst is produced by some earlier reaction $r_j \in R', j < i$, or is present in the food set $F$ \citep{Mossel:05}.

Note that the RAF in Fig. \ref{fig:example} is {\it not} a CAF. For example, molecule types $p_1$ and $p_2$ catalyze each other's formation, so one of the reactions $r_1$ and $r_2$ has to happen spontaneously (i.e., uncatalyzed) before the RAF set can come into existence. Of course chemical reactions can always happen without being catalyzed, as a catalyst primarily speeds up the {\it rate} at which that reaction happens. So, although the RAF set is present {\it topologically} (i.e., in the network representation), there may be a certain waiting time before it is realized {\it dynamically}. But once one of these two reactions happens spontaneously, the other reaction can proceed catalyzed and, in turn, create the necessary catalyst for the first reaction.

The example above is small enough to find the RAF set by eye. However, for larger networks this is clearly not convenient. Fortunately, there is an efficient {\it algorithm} for detecting RAF sets in arbitrary reaction networks.

\section{The RAF algorithm}

Given a CRS $Q=\{X,R,C,F\}$, an efficient (polynomial-time) algorithm exists for deciding whether $Q$ contains a RAF set or not. It is presented formally in Algorithm \ref{algo:RAF}, and roughly works as follows. Starting with the full set of reactions $R'=R$, the algorithm repeatedly calculates the closure of the food set relative to the current reaction set $R'$, and then removes from $R'$ all reactions that have none of their catalysts or not all of their reactants in this closure. This is repeated until no more reactions can be removed. If upon termination of the algorithm $R'$ is non-empty, then $R'$ is the unique {\it maximal} RAF set (maxRAF) contained in $Q$, i.e., a RAF that contains every other RAF in $Q$ as a subset \citep{Hordijk:04,Hordijk:15}. If upon termination $R'$ is empty, then $Q$ does not contain a RAF set.

\begin{algorithm}
\begin{algorithmic}
  \STATE{$R' = R$}
  \STATE{{\it change} = {\tt true}}
  \WHILE{({\it change})}
    \STATE{{\it change} = {\tt false}}
    \STATE{$\c_{R'}(F)$ = ComputeClosure ($F$, $R'$)}
    \FORALL{($r = (A,B) \in R'$)}
      \IF{($\nexists x \in \c_{R'}(F): (x,r) \in C \vee A \not\subseteq \c_{R'}(F)$)}
        \STATE{$R' = R' \setminus \{r\}$}
        \STATE{{\it change} = {\tt true}}
      \ENDIF
    \ENDFOR
  \ENDWHILE    
  \STATE{Return $R'$}  
\end{algorithmic}
\caption{RAF ($X$, $R$, $C$, $F$)}
\label{algo:RAF}
\end{algorithm}

Computing the closure of the food set relative to the current reaction set $R'$ is computationally the most expensive step in the RAF algorithm. It is presented formally in Algorithm \ref{algo:ComputeClosure}. Note that this algorithm is equivalent to the ``network expansion'' algorithm of \citet{Goldford:17}. However, the closure computation algorithm was already introduced  much earlier \citep{Hordijk:04}.

\begin{algorithm}
\begin{algorithmic}
  \STATE{$W = F$}
  \STATE{{\it change} = {\tt true}}
  \WHILE{({\it change})}
    \STATE{{\it change} = {\tt false}}
    \FORALL{($r = (A,B) \in R'$)}
      \IF{$(A \subseteq W \wedge B \not\subseteq W)$}
        \STATE{$W = W \cup B$}
        \STATE{{\it change} = {\tt true}}
      \ENDIF
    \ENDFOR
  \ENDWHILE
  \STATE{Return $W$}  
\end{algorithmic}
\caption{ComputeClosure ($F$, $R'$)}
\label{algo:ComputeClosure}
\end{algorithm}

A straightforward computational complexity analysis of the RAF algorithm gives a worst-case running time of $\O(|X||R|^3)$. However, with some additional book-keeping, such as keeping track of all reactions that each molecule is involved in, this can be reduced somewhat. In practice, the {\it average} running time on a simple polymer-based model of CRSs turns out to be sub-quadratic \citep{Hordijk:04}.

When the (full) network of Fig. \ref{fig:example} is given as input to the RAF algorithm, it will return the set $R' = \{r_1,r_2,r_3,r_4\}$ (i.e., the subset within the blue outline), which is the maxRAF for this network. However, note that this maxRAF contains two (nested) RAF subsets (subRAFs): $R_1'=\{r_1,r_2,r_4\}$ and $R_2'=\{r_1,r_2\}$. The subRAF $R_2'$ is a so-called {\it irreducible} RAF (iRAF), as it cannot be reduced any further without losing the RAF property. Such subRAFs and iRAFs can be detected with repeated applications of the RAF algorithm to the maxRAF after removal of one or more (random) reactions. Note that the maxRAF together with all its subRAFs form a {\it partially ordered set} \citep{Hordijk:12}.

Finally, the closure computation algorithm can be easily adjusted to find CAF sets. This is presented in Algorithm \ref{algo:CAF}. In this case, rather than checking whether all of a reaction's reactants are already included in $W$, we need to check whether all of its reactants {\it and} at least one catalyst is included in $W$. In addition, we need to keep track of which reactions were used to expand the set $W$. The set $S$ returned by the algorithm, if non-empty, is the unique (maximal) CAF set contained in $Q$, i.e., a CAF that contains every other CAF in $Q$ as a subset. Otherwise there is no CAF set.

\begin{algorithm}
\begin{algorithmic}
  \STATE{$W = F$}
  \STATE{$S = \emptyset$}
  \STATE{{\it change} = {\tt true}}
  \WHILE{({\it change})}
    \STATE{{\it change} = {\tt false}}
    \FORALL{($r = (A,B) \in R$)}
      \IF{$(A \subseteq W \wedge \exists x \in W: (x,r) \in C \wedge r \not\in S)$}
        \STATE{$W = W \cup B$}
        \STATE{$S = S \cup \{r\}$}
        \STATE{{\it change} = {\tt true}}
      \ENDIF
    \ENDFOR
  \ENDWHILE
  \STATE{Return $S$}  
\end{algorithmic}
\caption{CAF ($X$, $R$, $C$, $F$)}
\label{algo:CAF}
\end{algorithm}

\ \\
An implementation of the RAF algorithm (and several variants) is available in the program {\tt CatlyNet}, which runs on all main platforms (Windows, MacOS, and Linux) and includes a graphical user interface and some examples. It can be downloaded and installed for free from its \href{https://github.com/husonlab/catlynet}{\underline{GitHub repository}}. The program is described, together with examples of applications to biochemical networks, in \citet{Steel:20}.

\bibliographystyle{model5-names}\biboptions{authoryear}
\bibliography{methods}

\end{document}